\documentclass[amsmath,amssymb,superscriptaddress,nobalancelastpage,prx,twocolumn,aps]{revtex4-1}

\usepackage{graphicx}
\usepackage{varioref}
\usepackage{xr-hyper}
\usepackage{xcolor}
\usepackage{hyperref}
\usepackage{ulem}

\newcommand{\lico}{Li$_2$CuO$_2$}

\begin{document}

\title{Probing inter- and intrachain Zhang-Rice excitons in \lico\ and determining their binding energy}

\author{Claude Monney}\email{monney@physik.uzh.ch}
\affiliation{Research Department Synchrotron Radiation and Nanotechnology, Paul Scherrer Institut, 5232 Villigen PSI, Switzerland}
\affiliation{Institute of Physics, University of Zurich, Winterthurerstrasse 190, 8057 Zurich, Switzerland}
\author{Valentina Bisogni}
\affiliation{Research Department Synchrotron Radiation and Nanotechnology, Paul Scherrer Institut, 5232 Villigen PSI, Switzerland}
\affiliation{National Synchrotron Light Source II, Brookhaven National Laboratory, Upton, NY 11973, USA}
\affiliation{Leibniz Institute for Solid State and Materials Research, Helmholtzstrasse 20, 01171 Dresden, Germany}
\author{Ke-Jin Zhou}
\affiliation{Research Department Synchrotron Radiation and Nanotechnology, Paul Scherrer Institut, 5232 Villigen PSI, Switzerland}
\affiliation{Diamond Light Source, Harwell Science and Innovation Campus, Didcot, Oxfordshire, OX11 0DE, United Kingdom}
\author{Roberto Kraus}
\affiliation{Leibniz Institute for Solid State and Materials Research, Helmholtzstrasse 20, 01171 Dresden, Germany}
\author{Vladimir Strocov}
\affiliation{Research Department Synchrotron Radiation and Nanotechnology, Paul Scherrer Institut, 5232 Villigen PSI, Switzerland}
\author{G\"unter Behr\footnote{Deceased}}
\affiliation{Leibniz Institute for Solid State and Materials Research, Helmholtzstrasse 20, 01171 Dresden, Germany}
\author{Stefan-Ludwig Drechsler}
\affiliation{Leibniz Institute for Solid State and Materials Research, Helmholtzstrasse 20, 01171 Dresden, Germany}
\author{Helge Rosner} 
\affiliation{Max-Planck Institute for Chemical Physics of Solids, 01187 Dresden, Germany}
\author{Steve Johnston}
\affiliation{Department of Physics and Astronomy, The University of Tennessee, Knoxville, TN 37996, USA}
\affiliation{Joint Institute for Advanced Materials, The University of Tennessee, Knoxville, Tennessee 37996, USA}
\author{Jochen Geck}
\affiliation{Chemistry and Physics of Materials, Paris Lodron University Salzburg, 5020 Salzburg, Austria}
\author{Thorsten Schmitt}\email{thorsten.schmitt@psi.ch}
\affiliation{Research Department Synchrotron Radiation and Nanotechnology, Paul Scherrer Institut, 5232 Villigen PSI, Switzerland}

\begin{abstract}
Cuprate materials, like those hosting high temperature superconductivity,
represent a famous class of materials where the correlations between the
strongly entangled charges and spins produce complex phase diagrams. 
Several years ago the Zhang-Rice singlet 
was proposed as a natural quasiparticle in hole-doped cuprates.  
The occurance and binding energy of this quasiparticle, consisting of 
a pair of bound holes with antiparallel spins on the same CuO$_4$ plaquette,  
depends on the local electronic interactions, which are fundamental quantities 
for understanding the physics of the cuprates. Here, 
we employ state-of-the-art Resonant Inelastic X-ray
Scattering (RIXS) to probe the correlated physics of the CuO$_4$ plaquettes 
in the quasi-one dimensional chain cuprate \lico. By tuning the incoming photon
energy to the O $K$-edge, we populate bound states related to the Zhang-Rice 
quasiparticles in the RIXS process. Both intra- and interchain Zhang-Rice
singlets are observed and their occurrence is shown to depend on the
nearest-neighbor spin-spin correlations, which are readily probed in this
experiment. We also extract the binding energy of
the Zhang-Rice singlet and identify the Zhang-Rice triplet excitation in the RIXS spectra.
\end{abstract}

\maketitle


\section{Introduction}

As originally proposed by Zhang and Rice \cite{ZhangOrig}, a bound state formed
by two holes on the same plaquette is a natural quasiparticle in hole-doped
copper oxides. Called a Zhang-Rice (ZR) singlet in the case of opposite spins,
this quasiparticle consists of a singlet pair of holes, where one is localized
on the Cu$^{2+}$ ion and the other is delocalized on surrounding ligand
oxygens.  
In the undoped two-dimensional 
cuprates, the ZR singlet appears upon removing an electron from the
CuO$_4$ plaquette \cite{TjengARPES}. In the doped case,  
recent experimental results have
demonstrated that the ZR singlet remains stable in the ground state 
at all doping levels across the superconducting dome
up to the metallic overdoped regime \cite{Brookes2015}. These findings confirm 
that the ZR singlet picture has a relevant role in the description of the
electronic properties of the high-T$_c$ cuprates at all dopings \cite{Chen2013}. 
In addition, ZR singlet excitations have been
observed in several one-dimensional corner-sharing cuprates by using
electron spectroscopies \cite{TjengARPES,Neudert1998}. 

The existence of the ZRS has been also confirmed in the edge-sharing cuprate systems \cite{MatiksLCV,AtzkernEELS2000,MatiksPhD}, with the use of RIXS \cite{LearmonthRIXS,Kim2004,Duda2000,Vernay2008} among other techniques. In particular, the ZR singlet has been predicted in these materials \cite{MalekOptics,OkadaZRScorner}, especially in the
final state of the RIXS process \cite{OkadaZRSedge}, and has recently been
observed in \lico\ and CuGeO$_3$ \cite{MonneyPRL2012}. 

There are two important energy scales for the creation of ZR excitations
observed in undoped cuprate materials by RIXS: first, a hole has to be excited
from the Cu site to an O site, which involves the charge transfer energy $\Delta$.
Second, the excited hole on the O site must bind to the hole on the central Cu
site with a binding energy $B_{ZR}$. It is of great
relevance to have direct experimental access to these quantities, as they are fundamental to the physics of correlated
materials based on CuO$_4$ plaquettes.

We have chosen \lico\ for investigating the physics of ZR quasiparticles in a
simplified quasi-one-dimensional material. This prototypical realization of the
edge-sharing chain cuprates is made out of CuO$_4$ plaquettes without apical
oxygens.  Despite its low dimensionality, this system develops long range
magnetic order below $T_{M}=9$~K, where collinear ferromagnetic (FM) intrachain 
correlations  
and alternating antiferromagnetic (AFM) interchain correlations are thought to
be realized \cite{LorenzINS}.  

Here, we investigate the electronic excitations of \lico\ with RIXS and X-ray
absorption spectroscopy performed at the O $K$-edge
\cite{AmentReview,OkadaZRSedge}, revealing rich spectra showing peculiar charge
transfer excitations. Distinct excitonic ZR singlet
excitations are observed and are attributed to both intra- and interchain
excitations. Their nature is further investigated by means of temperature
dependent RIXS measurements. This method allows us to
probe not only the nearest-neighbor (nn) intrachain magnetic correlations, but also
the nn interchain magnetic correlations. 
Moreover, we also extract the ZR binding energy, $B_{ZR}$, from the energy loss position of
the ZR singlet and the charge transfer energy $\Delta$, 
which permits us to identify the ZR triplet excitonic excitation in the RIXS spectra.
Finally, our results confirm the intrachain long-range ferromagnetic order and
the interchain antiferromagnetic order in this material, emphasizing the
capability of RIXS in probing short range magnetic correlations.

\section{experimental and theoretical details}

\subsection{RIXS experiment} 
Experiments were performed at the ADRESS beamline \cite{beamline} of the Swiss
Light Source, Paul Scherrer Institut, using the SAXES spectrometer
\cite{SAXES}. RIXS spectra were typically recorded with a 2h acquisition time,
achieving  statistics of 100-150 photons on the peaks of interest (see below).
A scattering angle of $130^\circ$ was used and all the spectra were measured at
the specular position (at an incidence angle of $65^\circ$), meaning that no
momentum is transferred from the photons to the system along the chain direction. All
spectra were acquired with $\sigma-$polarization of light. The combined energy
resolution was 60 meV at the O K-edge ($\hbar\omega_i\sim530$ eV). \lico\
single crystals \cite{BehrLCO} (which are hygroscopic crystals) were cleaved
in-situ at the pressure of about $5\cdot10^{-10}$ mbar and at 20 K, producing
mirrorlike surfaces. The surface was oriented along the (101) direction, so
that the CuO$_4$ plaquettes were 21$^\circ$ away from the surface. This means
that the electric field of $\sigma-$polarized light was lying partially out of
the plaquette plane, its main component in the plaquette being along the
$c-$direction.
All of the RIXS spectra presented here are normalized to the
acquisition time, if not stated differently.

\subsection{Density functional theory calculations}
Scalar-relativistic density functional theory (DFT) electronic structure
calculations were performed using the full-potential FPLO code \cite{fplo},
version fplo9.01-35. 
The parametrization of Perdew-Wang \cite{PW} was
chosen for the exchange-correlation potential within the local
density approximtion (LDA).  
The calculations were carried out on a well converged mesh of 1152
$k$-points in the first Brillouin zone to ensure a high accuracy for details in the
electronic density of states. For the LDA$+U$ calculations, the
around-mean-field (AMF) double counting correction was applied. The
calculations were carried out using the experimental crystal structure
\cite{sapina1990}. 
The value of $U_{3d}$ used in the DFT calculations differs 
from the respective parameter $U_{dd}$ in the cluster model: the $U_{3d}$
is applied to all Cu-$3d$ basis states in the DFT calculations, 
whereas the $U_{dd}$ acts on the 3$d_{xy}$ states in the core 
Hamiltonian of the $pd$-Hubbard model. Naturally, this requires a
considerably smaller $U_{3d}$-value for the DFT calculations (including a 
slight basis dependence therein).

\subsection{RIXS cluster calculations} 
XAS and RIXS calculations were carried out using the
Kramers-Heisenberg formalism \cite{AmentReview}, where the initial, intermediate, and 
final states are obtained from small cluster exact diagonalization (ED). The 
calculation details are similar to those given in Ref.~\cite{MonneyPRL2012}. 
In order to model the interchain coupling, we considered a cluster formed from 
the two CuO$_4$ plaquettes with open boundary conditions that were 
bridged by a single Li atom. 
The unoccupied Li orbitals substantially increased the size 
of the Hilbert space. Our cluster therefore contained a limited 
orbital basis consisting of a single Cu $3d_{x^2-y^2}$ orbital on each Cu site, the 
O $2p_\sigma$ orbital on each oxygen site, and an effective $2s$ state 
on the Li site. 
The intra-plaquette Hamiltonian includes nearest-neightbor Cu-O and O-O 
hopping with $|t_{pd}| = 1$ and $|t_{pp}| = 0.65$ eV , respectively. 
The inter-plaquette O-Li hopping was taken to be $|t_{ps}| = 1$ eV. 
The onsite energies for the Cu, O, and Li orbitals were $\epsilon_d = 0$, $\epsilon_\sigma = 3.8$, 
and $\epsilon_s = -2$ eV, respectively. The on-site Hubbard interactions for 
the Cu $3d$ and O $2p$ orbitals were also included with $U_{dd} = 8$ and 
$U_{pp} = 4.1$ eV. All parameters are given in hole language.

The initial and final states were obtained for this cluster by diagonalizing 
the problem in the $N = 6$ hole sector with three spin up and three spin 
down holes. The 
intermediate states were found by diagonalizing in the $N = 5$ hole sector, 
with the inclusion of an additional core-hole potential $U_Q = 4$ eV on the 
oxygen site where the core hole was created. 
Finally, the reader should note that we have observed significant finite size 
effects in our previous work \cite{MonneyPRL2012} for the RIXS spectra 
when we have performed calculations on smaller single-chain clusters. Therefore 
the RIXS spectra shown in the main text should be considered as a simple qualitative 
description of the system.

\section{Results and discussion}

\subsection{RIXS spectra} 
\begin{figure}
\centering
\includegraphics[width=8.5cm]{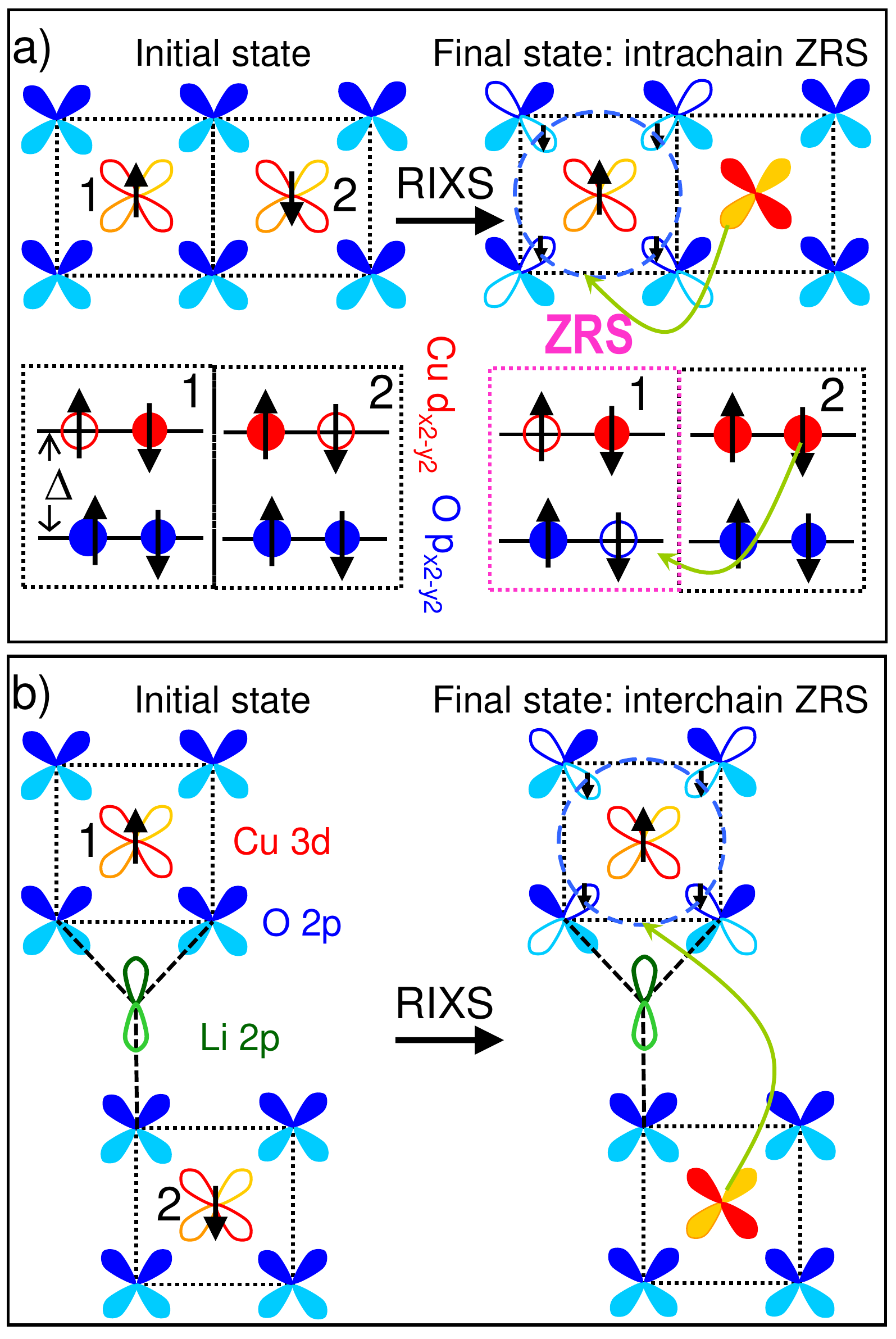}
\caption{\label{fig_1}
Schematic pictures of the creation of the ZR singlet (ZRS) excitons in the RIXS process. (a) Comparison of the RIXS initial and final states for the intrachain ZR singlet exciton, together with the corresponding energetic diagram. Full and empty symbols represent filled and empty states/orbitals. (b) A similar picture is shown for the case of the interchain ZR singlet exciton, for which the Li atom acts as a bridge between neighboring chains. The green arrow indicates the hole transfer between two plaquettes leading to excitonic excitations.
}
\end{figure}

During a RIXS experiment performed at the O $K$-edge, light excites directly an
electron from an O $1s$ core state into an O $2p$ valence state. However, due
to the hybridization of the O $2p$ states with the Cu $3d$ states in the
CuO$_4$ plaquettes, the excited electron can also fill a Cu $3d$ state. Such a
process can lead to the creation of an ZR exciton on a plaquette as a final
state. This is illustrated in Fig. \ref{fig_1} (a). Starting from an initial
state with two neighboring CuO$_4$ plaquettes on a single chain with both Cu in
a $d^9$ configuration, written in short $(d^9,d^9)$, a RIXS intermediate state
$(d^9,\underline{1s}\:d^{10})$ (with one hole in the O 1s states,
$\underline{1s}$) is reached, deexciting afterwards to a final state
$(d^9\underline{L},d^{10})$ ($\underline{L}$ denotes a ligand hole). Comparing the
initial to the final state, one sees that a non-local charge-transfer has
occurred from one plaquette to another \cite{DudaNiO}, leading to an excitonic
excitation due to the local charge imbalance \cite{MatiksLCV}. From the
conservation of spin in the process, an initial state with antiparallel spins
(panel (a)) on the neighboring plaquettes ends up in a final state with two
holes having antiparallel spins on the left plaquette, $d^9\underline{L}$,
which is a ZR singlet \cite{ZhangOrig}. It is straightforward to conclude that
an initial state with parallel spins on the neighboring plaquettes will give
rise to a ZR triplet.

In \lico\ and other spin-chain cuprates, there is another possibility for the
creation of a low-energy ZR singlet excitation, which takes place between two
neighboring CuO$_2$ chains, as illustrated in Fig. \ref{fig_1} (b). In \lico,
the donor Li atoms lie between such chains \cite{MizunoInterC,HoppeXTal} and
can act as a bridge for the electron created at the O site in the intermediate
state of the RIXS process.  This results in the creation of an interchain ZR
singlet exciton in the final state.

\begin{figure}
\centering
\includegraphics[width=9.0cm]{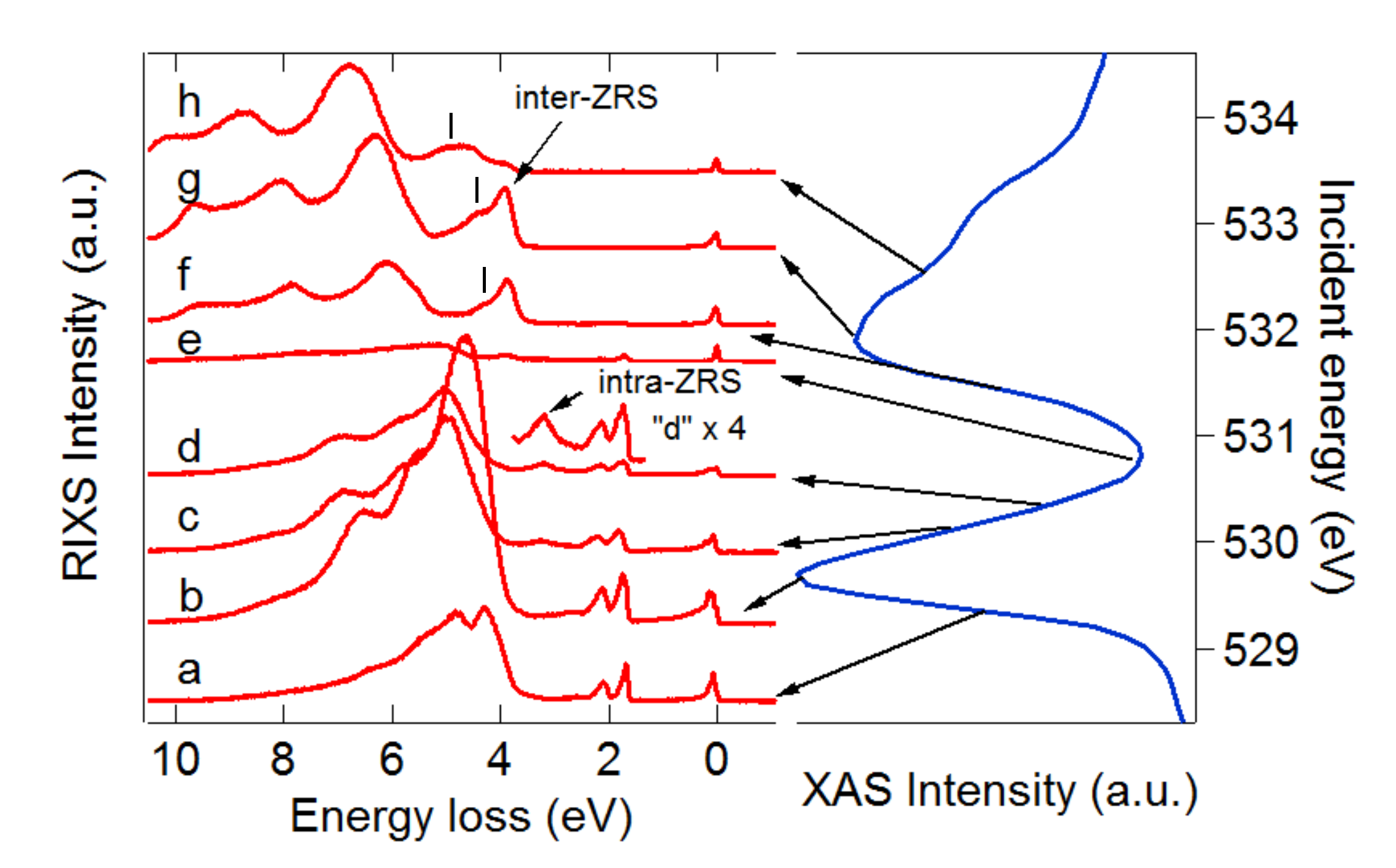}
\caption{\label{fig_2} Incident energy dependence of the RIXS spectra. (Right) XAS spectrum measured at the O $K$-edge on \lico\ with $\sigma-$polarized light and at 20 K, (left) together with the corresponding RIXS spectra (on an energy loss scale) measured at incident photon energies given by the arrows relative to the XAS spectrum energy scale. The position of the intra- and interchain ZR singlet excitations are emphasized by arrows. The position of the possible dispersive ZR fluorescence excitation is also shown by vertical lines. (see text). 
}
\end{figure}

In the first part of our experimental approach, we measured X-ray absorption
spectroscopy (XAS) at the O $K$ edge on \lico\ samples, in order to determine
the appropriate incident photon energy for RIXS.
In Fig. \ref{fig_2} (right), we show the XAS spectrum obtained at 20 K (which
is above $T_{M}$). It is composed of two main peaks. The peak at 529.7 eV is
due to the upper Hubbard band (UHB) consisting mainly of Cu $3d_{x^2-y^2}$
states hybridized with O $2p$ states of the same symmetry \cite{NeudertXAS}.
Next to this sharp line, there is another broad structure extending from about
$\hbar\omega_i=$~531 eV to 534 eV, with a maximum at 532 eV. This structure is
due to Li $2s/2p$ states hybridized with (in-chain) O states. 
This interpretation of the XAS features is verified by our density functional
theory (DFT) calculations shown in Fig. \ref{fig_3} (a) (see below). The character of the
low-energy unoccupied states and their energy scale compare well with our XAS
spectrum.

The final states of the XAS spectrum described above correspond to intermediate
states in the RIXS process \cite{AmentReview,RIXSReview}. Specifically, by
tuning the incoming photon energy along the XAS spectrum of \lico, as shown by
the arrows on Fig. \ref{fig_2} (right), we select which intermediate states are
involved in the RIXS process. 
The resulting O $K$-edge RIXS spectra measured also at 20 K are plotted in an
energy loss scale in Fig. \ref{fig_2} (left). At first sight, one recognizes
remarkably rich spectra, exhibiting different sharp peaks. 

The intense broad structure between 4 and 10 eV originates from higher-energy
charge transfer excitations. Most of this spectral weight disperses as a function of the incident photon energy above $\hbar\omega_i=$531
eV, identifying a significant amount of these excitations as fluorescence, coming from the
valence states that have a mixed Cu-O character according to first-principle
calculations \cite{WehtDFT}. More interesting are the peaks situated between
the elastic line (at 0 energy loss) and the higher-energy charge transfer
excitations.  Two sharp peaks appear at about 2 eV and do not move in energy
loss as a function of incident energy \--- they exhibit a so-called Raman
behavior. This behavior (and the involved energy scale, which is in rather good
agreement with quantum chemistry calculations \cite{HuangDDcalcs}) is typical
of local excitations occuring between the different $d$-orbitals of a single Cu
site, called $dd-$excitations.  This is supported by the fact that they
resonate when the incident energy is tuned to the UHB, as already observed by
Learmonth {\it et al.} \cite{LearmonthRIXS}. 

Our main interest relates to the Raman excitations observed at energy losses
between the $dd-$excitations and the higher-energy charge transfer excitations.
An interesting peak appears (in spectra $c,d$ of Fig. \ref{fig_2}) at about 3.2
eV energy loss when selecting incident photon energies corresponding to
exciting the O 1$s$ electron into the UHB (around $\hbar\omega_i=$530.1 eV). We
have already identified this RIXS peak as the intrachain ZR singlet exciton
\cite{MonneyPRL2012}. 
Furthermore, by tuning the incident photon energy to the Li-hybridized XAS
structure (around $\hbar\omega_i=$531.9 eV), we create RIXS-intermediate
states, where the O 1$s$ electron is excited into Li-hybridized states. In this
case, another Raman contribution appears (in spectra $f,g,h$ of Fig.
\ref{fig_2}) at about 4.0 eV energy loss. We will show now that it corresponds
to the interchain ZR singlet exciton.

\subsection{Calculations}
\begin{figure}[h]
\centering
\includegraphics[width=8.5cm]{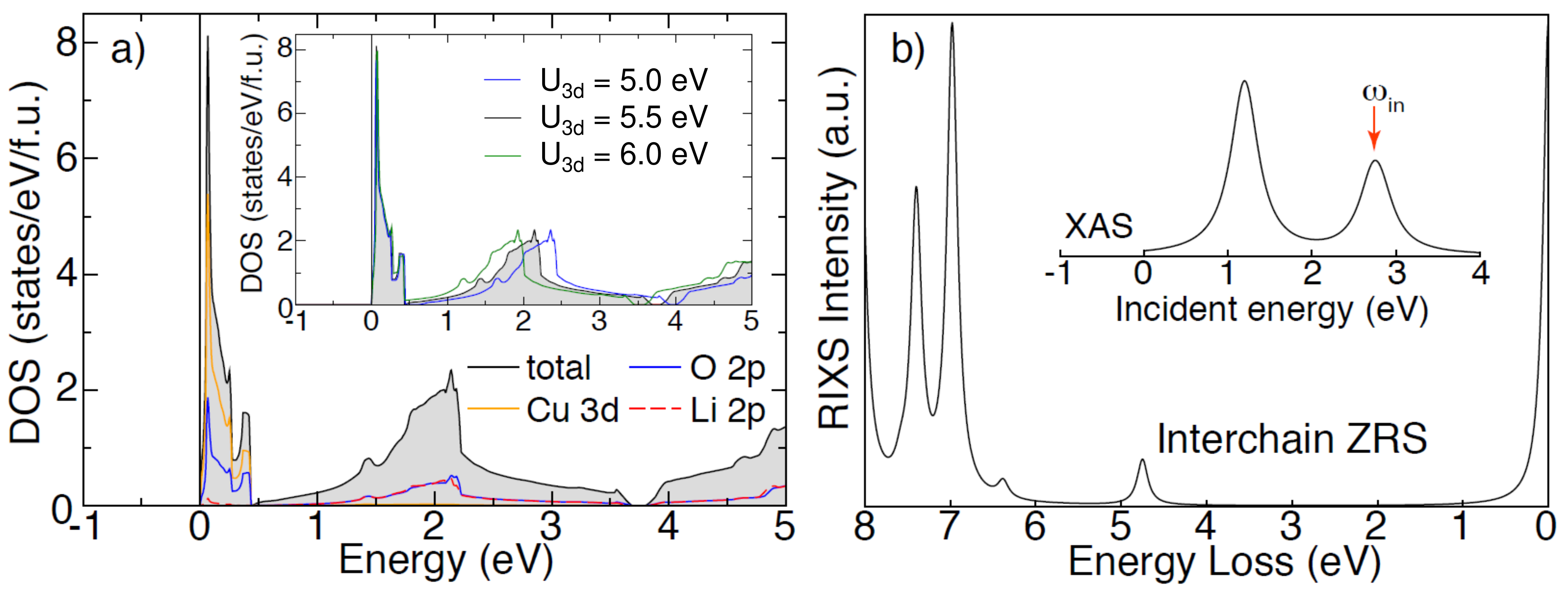}
\caption{\label{fig_3} 
DFT and RIXS cluster calculations. 
(a) Calculated DOS within DFT+$U$ for the unoccupied states of \lico. 
The main panel shows the orbital-resolved DOS of \lico\ for $U_{3d} = 5.5$ eV. 
The inset shows the variation in the total DOS for different 
values of $U_{3d}$ (the 0 eV reference energy is placed at the lower band edge of the unoccupied states). (b) 
Calculated XAS (inset) and RIXS for a ED cluster calculation based on a 2 CuO$_4$ plaquettes bridged by a Li atom.
The red arrow indicates at which incident energy the RIXS spectrum has been
calculated.
}
\end{figure}

To better understand the structure of the XAS spectrum we have performed 
DFT calculations to determine the nature of the unoccupied
states in \lico.  
In order to account for the strong electron-electron interaction in this
material, we introduced an orbital dependent Coulomb repulsion 
$U_{3d}$ at the Cu site, varying between 5.0 and 6.0 eV. This range 
for $U_{3d}$ is known empirically to describe well the magnetic properties
due to the Heisenberg exchange between the localized 
Cu-moments in many Cu-O compounds \cite{LorenzINS,schmitt2009,wolter2012}.
The calculated density of states (DOS) is shown in Fig. \ref{fig_3} (a) and is compared to the XAS spectrum in Fig. \ref{fig_2} as a first approximation.
With $U_{3d}=5.5$ eV, the DOS obtained here reproduces well the two peaks observed in the XAS (Fig. \ref{fig_2}, right). 
The partial DOS shown in Fig. \ref{fig_3} (a) confirms that the second peak (at higher energies) in the DOS is mostly due to O 2$p$ states hybridized with Li 2$p$, in agreement with a previous XAS study \cite{NeudertXAS}. This supports the idea that an electron is promoted into such Li $2p$ hybridized states in the intermediate state of RIXS when the incident energy is tuned at about 532 eV, i.e. 1.5 eV above the upper Hubbard band (see Fig. \ref{fig_2}).

We now show that an electron excited with an incident energy of about 532 eV
can exploit the unoccupied Li 2$p$ states to travel from one CuO$_2$
chain to the neighboring one. For this purpose, we have calculated RIXS
intensities at the O $K$-edge  using a small cluster ED calculation
based on a model system consisting of two CuO$_4$ plaquettes bridged by a Li
atom (as in Fig. \ref{fig_1} (b)). 
Such a dual-chain geometry involving many different electronic orbitals has
never been used for such calculations so far.  (For example, in our previous work \cite{MonneyPRL2012}, we have
simulated a single CuO$_2$ chain with up to 5 CuO$_4$ plaquettes.)

The ED results for the RIXS spectrum is shown in Fig. \ref{fig_3} (b), while   
the corresponding XAS spectrum is shown in the inset.
The XAS spectrum reproduces well the two main peaks of the measured XAS, given the
simplicity of the model system. The RIXS spectrum has been calculated for an
incident energy tuned to the second peak in the XAS (see the red arrow in the
inset in Fig. \ref{fig_3} (b)). In addition to the elastic line at 0 eV energy
loss, a series of charge transfer excitations appear above 6 eV energy
loss, in agreement with the experiment (see Fig. \ref{fig_2} (left)). Most
interestingly, the calculation displays another peak at about 4.8 eV energy
loss. This corresponds to a $d^9\underline{L}$ state, where a hole has been
transferred from one CuO$_2$ chain to the other during the RIXS process. This
excitation therefore corresponds to the interchain ZR singlet exciton, which is
excited when the incident energy is tuned to Li-O hybridized intermediate
states.  While the energy of this RIXS excitation is somewhat larger than what
is experimentally observed (4.0 eV) due to the simplicity of the model adopted
here, this calculation together with the DOS shown above, demonstrates that it
is possible to create an interchain ZR singlet exciton at the O $K$-edge in
\lico.

\subsection{Temperature dependence}
\begin{figure*}
\centering
\includegraphics[width=\textwidth]{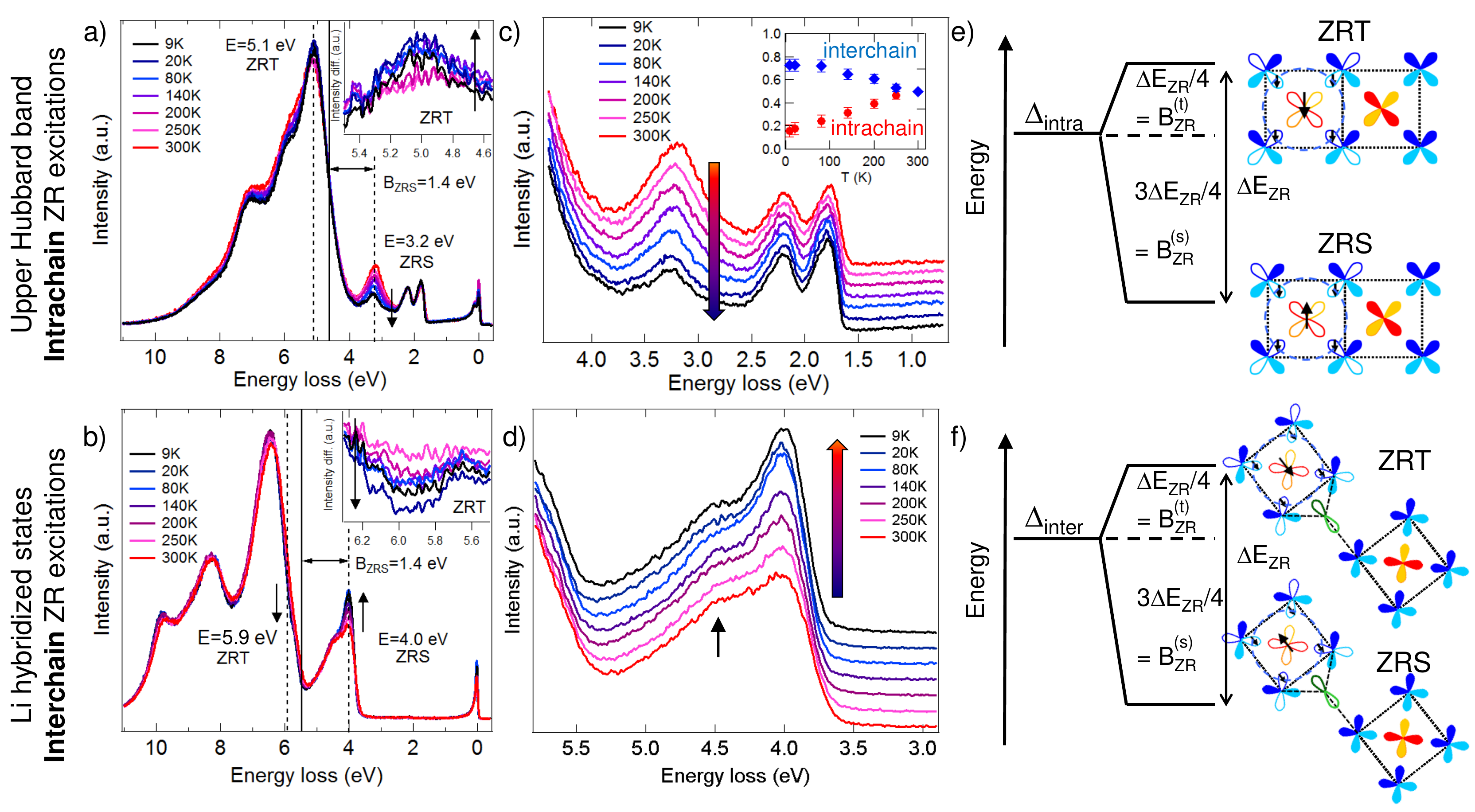}
\caption{\label{fig_4}
Temperature dependence of RIXS intensities of the intrachain and interchain ZR singlet excitations. RIXS spectra measured as a function of temperature at the O $K$-edge of \lico, with $\sigma$-polarization of light and an incident energy of (a) $\hbar\omega_i=$530.1 eV (on the edge of the UHB) and of (b) $\hbar\omega_i=$531.9 eV (on the Li-hybridized states).  The RIXS spectra in (a) are normalized to the total intensity area of the $dd$-excitations \cite{MonneyPRL2012}. In (a) and (b), the insets show difference RIXS spectra, for which the spectrum at 300 K has been subtracted and the vertical continuous line indicate the energy position of the charge transfer energies $\Delta_{inter,intra}$ (see text). (c) and (d): corresponding zoom on the ZR singlet excitons.  The spectra are shifted in intensity by a constant offset for better visibility. (c) The inset shows the integrated RIXS intensity (normalized to 0.5 at room temperature) of the ZR singlet excitons as a function of temperature. Schematic energy scale for the (e) intrachain and (f) interchain ZR excitons, depicting how the ZR singlet excitation energy is shifted to energy losses lower than the charge transfer energy, $\Delta$, by its binding energy, $B_{ZR}^{(s)}$, while the ZR triplet (ZRT) excitation energy is shifted above $\Delta$. The splitting between the ZR singlet and ZR triplet excitation energies is $\Delta E_{ZR}$.
}
\end{figure*}

The conservation of spin in this RIXS process allows us to determine the local
spin correlations by looking at the ZR exciton intensity in the RIXS data as a
function of temperature (see also Fig.  \ref{fig_1}) \cite{MonneyPRL2012}.
Having this in mind, we have measured the temperature dependence of the
RIXS intensity of both the intrachain and the interchain ZR singlet exciton peak at 3.2 eV and 4.0 eV energy loss, see Fig. \ref{fig_4} (a) and (b) respectively.
The spectra zoomimg on the intrachain and the  interchain ZR exciton peaks are shown in Fig. \ref{fig_4} (c) and (d), respectively.  The temperature dependence of these two Raman-like peaks is opposite, as summarized in the inset of Fig. \ref{fig_4} (c), which shows the integrated
peak intensities vs temperature: the interchain ZR singlet exciton is
increasing in intensity when temperature is lowered, while the intrachain
ZR singlet exciton is decreasing in intensity.

These temperature behaviors allow us now to relate the nature of these ZR
singlet excitons to the magnetic correlations. As mentioned above, while
long-range FM order occurs in the CuO$_2$ chains of \lico\ below the critical
temperature, $T_{M}=9$ K, long-range AFM order develops across the chains
\cite{LorenzINS}.  Above $T_{M}$, the magnetic order along the chains is
expected to vanish, transforming into short range order of decreasing
correlation length as temperature increases. As a consequence, parallel spins
are dominating the intrachain nearest neighbor spin correlations at low
temperatures, but their occurence decreases as temperature
increases. This is consistent with the temperature dependence of the intrachain
ZR singlet peak in Fig. \ref{fig_4} (c), the intensity of which decreases as the temperature decreases. 
Similarily, the development of interchain AFM order at low temperatures
increases the probability of finding antiparallel spins on neighboring chains. 
Thus, the intensity of the interchain ZR singlet increases 
as the temperature is decreased, as shown in Fig. \ref{fig_4}
(d).  More generally, this observation confirms that O $K$-edge RIXS is capable
of probing the short-range magnetic correlations of such low dimensional
systems.

\subsection{Singlets and triplets}
As expected from our schematic description in
Fig. \ref{fig_1}, we observe both the intra- and interchain ZR singlet excitons
in our RIXS data on \lico.  Comparing the energy loss of these peaks allows us
to gain substantial physical information on the ZR physics in this prototypical
low dimensional cuprate. First, following the processes depicted in Fig.
\ref{fig_1}, we estimate the energy cost of such an excitation, starting from a
$(d^9,d^{9})$ configuration (charge neutral). The transfer of a hole from a Cu
$d_{x^2-y^2}$ orbital to ligand $p_{x^2-y^2}$ orbitals of the neighboring
CuO$_4$ plaquette costs the charge transfer energy, $\Delta$. (We can neglect
the cost of the broken magnetic bonds in comparison to this energy scale in the case of
edge-shared geometry since the Cu-Cu exchange energies are less than 20
meV \cite{LorenzINS,MalekOptics}.)
As explained above, this RIXS final state is now
$(d^9\underline{L},d^{10})$, meaning that charge has been transferred from
one plaquette to another, leading to the creation of an electron-hole bound
state between two plaquettes forming an exciton. 
This excitonic
contribution to the energy cost is contained already in the Coulombic part of the charge transfer
energy \cite{OhtaCT}. As a consequence, we distinguish between the
interchain $\Delta_{inter}$ and intrachain $\Delta_{intra}$ charge transfer
energies, due to the different distances between the hole and the electron of
the exciton in these two processes. 
The energy loss of the ZR excitation in the RIXS process is given by
$E_{inter,intra}^{(s,t)}=\Delta_{inter,intra}-B_{ZR}^{(s,t)}$, with
$B_{ZR}^{(s,t)}$ being the binding energy of the ZR state, which can be singlet ($s$) 
or triplet ($t$). In other words, the binding energy of the ZR excitation is defined as
the energy difference between the charge transfer energy $\Delta$ and its excitation
energy (which is smaller than $\Delta$). We define additionally the
energy splitting between the ZR singlet and ZR triplet excitons as $\Delta
E_{ZR}=B_{ZR}^{(t)}-B_{ZR}^{(s)}$.

We recently evaluated the intrachain charge transfer energy in \lico\ to be
$\Delta_{intra}=4.6$ eV by comparing RIXS spectra with calculations done on a multiplaquette CuO$_2$ 
chain \cite{JohnstonSub}. Together with the excitation energy
of the intrachain ZR singlet, $E_{intra}^{(s)}=3.2$ eV, this gives us a ZR singlet binding
energy of $B_{ZR}^{(s)}=1.4$ eV in \lico. This is a first important result.
This ZR singlet binding energy must be the same for the interchain ZR singlet, since the
plaquettes where the ZR singlet takes place in the final state of the RIXS process,
are the same. From the excitation energy of the interchain ZR singlet,
$E_{inter}^{(s)}=4.0$ eV, we infer then an interchain charge transfer energy of
$\Delta_{inter}=5.4$ eV, which is 0.8 eV higher than the intrachain one. This energy difference
is probably coming from the different contribution of the nearest-neighbor
Coulomb interaction $U_{pd}$ in the final state of the intrachain vs. interchain
charge transfer (i.e. in the excitonic contribution).

Having identified intrachain and interchain ZR singlet, we now turn to the ZR triplet
excitations. For this purpose, we show in Fig. \ref{fig_4} the RIXS spectra of
\lico\ measured on a 10 eV range for incident energies of (a)
$\hbar\omega_i=$530.1 eV and (b) $\hbar\omega_i=$531.9 eV corresponding to
exciting intrachain and interchain ZR excitons, respectively. 
For $B_{ZR}^{(s)}=1.4$ eV, we expect the spin singlet to lie at an energy
$\Delta-3\Delta E_{ZR}/4$ and the spin triplet at an energy
$\Delta+1\Delta E_{ZR}/4$ with respect to the charge transfer energy, as
illustrated schematically in Fig. \ref{fig_4} (e) and (f). Additionally, the
ZR triplet exciton must have the inverse temperature behavior as that of the ZR singlet.
Consequently, the ZR triplet should be located at about $B_{ZR}^{(t)}=0.47$ eV above the charge
transfer energy, involving an energy separation $\Delta
E_{ZR}=1.87$ eV between the ZR singlet and the ZR triplet  
\footnote{It is important to stress that the energy loss position of the ZR
triplet excitation is relatively insensitive to the uncertainty on the value of
$\Delta_{inter,intra}$.}. This energy scale has been
confirmed from the singlet/triplet splitting obtained for two holes on a
single CuO$_4$ plaquette with open boundary conditions (not shown here).
In the case of the intrachain ZR excitons, the ZR triplet should be located at about
5.1 eV energy loss. At this energy in Fig. \ref{fig_4} (a) (vertical dashed
line), we see a large peak, the intensity of which has a temperature behavior
opposite to that of the intrachain ZR singlet (vertical arrows). We identify this
excitation as the intrachain ZR triplet exciton. In a similar way, we expect to see
the interchain ZR triplet exciton at about 5.9 eV energy loss. In the RIXS spectrum of
Fig. \ref{fig_4} (b), we do not distinguish a clear peak, but closer
inspection (see inset) reveals a small shoulder, the intensity of which is decreasing with
decreasing temperature, as expected for an excitation suppressed by interchain
antiferromagnetic correlations. We tentatively identify this excitation as the
interchain ZR triplet exciton.
More generally, our analysis shows that the identification in RIXS data of the ZR singlet and ZR triplet excitons permits us to directly extract $\Delta
E_{ZR}$ from their energy splitting. This, in turn, delivers the value of the ZR binding energies, $B_{ZR}^{(s,t)}$, as shown schematically in Fig. \ref{fig_4} (e) and (f).

Finally, we comment on the work of Learmonth \textit{et al.}
\cite{LearmonthRIXS} in the light of our new results. There, the authors
tentatively attributed to a ZR triplet excitation a shoulder at 4.1 eV energy loss near the fluorescence, resonating on a large incident energy range around the UHB.
Here, we identify this excitation as the interchain ZR singlet excitation, as it 
clearly resonates at higher incident energies. We attribute this
discrepancy in the interpretation to the higher energy resolution and
statistics of our experiment,as well as the fact that no temperature dependent study could be performed in Ref. \cite{LearmonthRIXS}. In Ref. \cite{LearmonthRIXS} the
intrachain ZR singlet was not observed, as the corresponding incident energy (slightly
detuned from the UHB) was not used.
Interestingly, a ZR singlet fluorescence excitation was proposed for describing a
fluorescence-like excitation developing at incident energies tuned to the Li-O
hybridized states in the XAS. Due to our higher energy resolution, we can distinguish here
this possible excitation (located by an arrow in Fig. \ref{fig_4} (d), at about 4.5 eV, and by vertical lines in Fig. \ref{fig_2} (left))
from the interchain ZR singlet exciton. Furthermore, it does not display any
temperature dependence, confirming Learmonth \textit{et al.}'s assumption
\cite{LearmonthRIXS}.

\section{Conclusions}
We have performed Resonant Inelastic X-ray Scattering measurements at the O
$K$-edge on the edge-sharing chain cuprate, \lico. Rich RIXS spectra are
observed with specific charge transfer excitations, which are understood as
Zhang-Rice singlet and triplet excitons created in the final state of the RIXS
process. By analysing the character of the states involved in the final states
of the X-ray absorption spectra and thus in the intermediate states of the RIXS
spectra, we identify interchain ZR exciton excitations. These are confirmed by
RIXS cluster calculations.
Both the intrachain and interchain ZR excitations are measured as a function of
temperature and their strong temperature dependent behavior is intimately
related to intrachain and interchain nearest neighbor magnetic correlations. This permit us, using RIXS, to confirm in \lico\ the development of both intrachain ferromagnetic order and  interchain antiferromagnetic order at low temperature.
The corresponding ZR triplet excitons are also observed in the RIXS spectra.
With this work, we demonstrate how it is possible to estimate several fundamental quantities including the ZR singlet 
binding energy, as well as interchain and intrachain charge transfer energies, from the energy loss position of these excitonic excitations.

\section{acknowledgments}
We acknowledge fruitful discussions with H.M. R\o nnow, U. Staub, J. van der Brink, J. M{\'a}lek, R. Kuzian and K. Wohlfeld. S.L.D. thanks A.Boris and D.Efremov for discussions of the role of interchain ZR excitons.
The experimental part of this work was performed at the ADRESS beamline of the Swiss Light Source at the Paul Scherrer Institut, Switzerland.
This project was supported by the Swiss National Science Foundation and its National Centre of Competence in Research MaNEP.
This research has been jointly funded by the German Science Foundation and the Swiss National Science Foundation within the D-A-CH program
(SNSF Grant No. $200021L\_141325$ and DFG Grant No. GE $1647/3-1$).
C.M. gratefully acknowledges the support by the SNSF under grant $PZ00P2\_ 154867$. 
S.J. is supported by the University of Tennessee’s Science Alliance Joint Directed Research and Development (JDRD) program, a collaboration with Oak Ridge National Laboratory.
J.G. gratefully acknowledges the support by the Collaborative Research Center SFB 1143.


\begin{thebibliography}{99}

\bibitem{ZhangOrig}
F.C. Zhang and T.M. Rice, \textit{Effective Hamiltonian for the superconducting Cu oxides}, Phys. Rev. B {\bf 37}, 3759 (1988).

\bibitem{TjengARPES}
L.H. Tjeng, B. Sinkovic, N.B. Brookes, J.B. Goedkoop, R. Hesper, E. Pellegrin, F.M.F. de Groot, S. Altieri, S.L. Hulbert, E. Shekel and G.A. Sawatzky, \textit{Spin-Resolved photoemission on anti-ferromagnets: Direct observation of Zhang-Rice singlets in CuO}, Phys. Rev. Lett. {\bf 78}, 1126 (1997).

\bibitem{Brookes2015}
N. B. Brookes, G. Ghiringhelli, A.-M. Charvet, A. Fujimori, T. Kakeshita, H. Eisaki, S. Uchida, and T. Mizokawa, \textit{Stability of the Zhang-Rice singlet with doping in Lanthanum Strontium Copper oxide across the superconducting dome and above}, Phys. Rev. Lett. {\bf 115}, 027002 (2015).

\bibitem{Chen2013}
C.-C. Chen, M. Sentef, Y. F. Kung, C. J. Jia, R. Thomale, B. Moritz, A. P. Kampf and T. P. Devereaux, \textit{Doping evolution of the oxygen K-edge x-ray absorption spectra of cuprate superconductors using a three-orbital Hubbard model}, Phys. Rev. B {\bf 87}, 165144 (2013).

\bibitem{Neudert1998}
R. Neudert, M. Knupfer, M.S. Golden, J. Fink, W. Stephan, K. Penc, N. Motoyama, H. Eisaki and S. Uchida, \textit{Manifestation of spin-charge separation in the dynamic dielectric response of one-dimensional Sr$_2$CuO$_3$}, Phys. Rev. Lett. {\bf 81}, 657 (1998).

\bibitem{MatiksLCV}
Y. Matiks, P. Horsch, R.K. Kremer, B. Keimer and A.V. Boris, \textit{Exciton doublet in the Mott-Hubbard insulator LiCuVO$_4$ identified by spectral ellipsometry}, Phys. Rev. Lett. {\bf 103}, 187401 (2009).

\bibitem{AtzkernEELS2000}
S. Atzkern, M. Knupfer, M. S. Golden, J. Fink, C. Waidacher, J. Richter, and K. W. Becker, N. Motoyama, H. Eisaki, and S. Uchida, \textit{Dynamics of a hole in a CuO$_4$ plaquette: Electron energy-loss spectroscopy of Li$_2$CuO$_2$}, Phys. Rev. B {\bf 62}, 7845 (2000).

\bibitem{MatiksPhD}
Y. Matiks, \textit{Spectroscopic ellipsometry of spin-chain cuprates and LaNiO$_3$-based heterostructures}, PhD thesis, University of Stuttgart, 2011.

\bibitem{LearmonthRIXS}
T. Learmonth, C McGuinness, P.-A. Glans, J.E. Downes, T. Schmitt, L.-C. Duda, J.-H. Guo, F.C. Chou and K.E. Smith, \textit{Observation of multiple Zhang-Rice excitations in a correlated solid: Resonant inelastic X-ray scattering study of Li$_2$CuO$_2$}, Euro. Phys. Lett. {\bf 79}, 47012 (2007).

\bibitem{Kim2004}
Y.-J. Kim, J. P. Hill, F. C. Chou, D. Casa, T. Gog, and C. T. Venkataraman, \textit{Charge and orbital excitations in Li$_2$CuO$_2$}, Phys. Rev. B {\bf 69}, 155105 (2004).

\bibitem{Duda2000} L.-C. Duda, J. Downes and C. McGuinness, T. Schmitt and A. Augustsson, K. E. Smith, G. Dhalenne and A. Revcolevschi, \textit{ Bandlike and excitonic states of oxygen in CuGeO3: Observation using polarized resonant soft-x-ray emission spectroscopy}, Phys. Rev. B {\bf 61}, 4186 (2000).

\bibitem{Vernay2008} F. Vernay, B. Moritz, I. S. Elfimov, J. Geck, D. Hawthorn, T. P. Devereaux and G. A. Sawatzky, \textit{ Cu K-edge resonant inelastic x-ray scattering in edge-sharing cuprates }, Phys. Rev. B {\bf 77}, 104519 (2008).

\bibitem{OkadaZRScorner}
K. Okada and A. Kotani, \textit{Copper-related information from the oxygen 1s resonant x-ray emission in low-dimensional cuprates}, Phys. Rev. B {\bf 65}, 144530 (2002).

\bibitem{MalekOptics}
J. M{\'a}lek, S.-L. Drechsler, U. Nitzsche, H. Rosner and H. Eschrig, \textit{Temperature-dependent optical conductivity of undoped cuprates with weak exchange}, Phys. Rev. B {\bf 78}, 060508(R) (2008).

\bibitem{OkadaZRSedge}
K. Okada and A. Kotani, \textit{Zhang-Rice singlet-state formation by oxygen 1s resonant x-ray emission in edge-sharing copper-oxide systems}, Phys. Rev. B {\bf 63}, 045103 (2001).

\bibitem{MonneyPRL2012}
C. Monney, V. Bisogni, K.J. Zhou, R. Kraus, V.N. Strocov, G. Behr, J. M{\'a}lek, R. Kuzian, S.-L. Drechsler, S. Johnston, A. Revcolevschi, B. Buchner, H.M. Ronnow, J. van den Brink, J. Geck, T. Schmitt, \textit{Determining the short-range spin correlations in the spin-chain Li$_2$CuO$_2$ and CuGeO$_3$ compounds using resonant inelastic x-ray scattering}, Phys. Rev. Lett. {\bf 110}, 087403 (2013).

\bibitem{LorenzINS}
W.E.A. Lorenz, R.O. Kuzian, S.-L. Drechsler, W.-D. Stein, N. Wizent, G. Behr, J. M{\'a}lek, U. Nitzsche, H. Rosner, A. Hiess, W. Schmidt, R. Klingeler, M. Loewenhaupt and B. Büchner, \textit{Highly dispersive spin excitations in the chain cuprate Li$_2$CuO$_2$}, Euro. Phys. Lett. {\bf 88}, 37002 (2009). 

\bibitem{AmentReview}
L. J. P. Ament, M. van Veenendaal, T. P. Devereaux, J. P. Hill, J. van den Brink, \textit{Resonant inelastic x-ray scattering studies of elementary excitations}, Rev. Mod. Phys. {\bf 83} 705 (2011).

\bibitem{beamline}
V.N. Strocov, T. Schmitt, U. Flechsig, T. Schmidt, A. Imhof, Q. Chen, J. Raabe, R. Betemps, D. Zimoch, J. Krempasky, X. Wang, M. Grioni, A. Piazzalunga and L. Patthey, \textit{High-resolution soft x-ray beamline ADRESS at the Swiss Light Source for resonant inelastic x-ray scattering and angle-resolved photoelectron spectroscopies}, J. Synchrotron Rad. {\bf 17}, 631 (2010).

\bibitem{SAXES}
G. Ghiringhelli, A. Piazzalunga, C. Dallera, G. Trezzi, L. Braicovich, T. Schmitt, V.N. Strocov, R. Betemps, L. Patthey, X. Wang and M. Grioni, \textit{SAXES, a high resolution spectrometer for resonant x-ray emission in the 400–1600 eV energy range}, Rev. Sci. Instrum. {\bf 77}, 113108 (2006).

\bibitem{BehrLCO}
N. Wizent, G. Behr, W. L\"oser, B. B\"uchner and R. Klingeler, \textit{Challenges in the crystal growth of Li$_2$CuO$_2$ and LiMnPO$_4$}, J. Cryst. Growth {\bf 318}, 995 (2011).

\bibitem{fplo} 
K. Koepernik, and H. Eschrig, \textit{Full-potential nonorthogonal local-orbital minimum-basis band-structure scheme}, Phys. Rev. B 59, 1743 (1999); I. Opahle, K. Koepernik, and H. Eschrig, \textit{Full-potential band-structure calculation of iron pyrite}, Phys. Rev. B {\bf 60}, 14035 (1999); http://www.fplo.de.

\bibitem{PW}
J.P. Perdew and Y. Wang, \textit{Accurate and simple analytic representation of the electron-gas correlation energy}, Phys. Rev. B {\bf 45}, 13244 (1992).

\bibitem{sapina1990} 
F. Sapina, J. Rodriguez Carvajal, M.J. Sanchis, R. Ibanez, A. Beltran, and D. Beltran, \textit{Crystal and magnetic structure of Li$_2$CuO$_2$}, Solid State Communications {\bf 74}, 779-784 (1990).

\bibitem{DudaNiO}
L.-C. Duda, T. Schmitt, M. Magnuson, J. Forsberg, A. Olsson, J. Nordgren, K. Okada and A. Kotani, \textit{Resonant inelastic x-ray scattering at the oxygen K resonance of NiO:
Nonlocal charge transfer and double-singlet excitations}, Phys. Rev. Lett. {\bf 96}, 067402 (2006).

\bibitem{MizunoInterC}
Y. Mizuno, T. Tohyama and S Maekawa, \textit{Interchain interactions and magnetic properties of Li$_2$CuO$_2$}, Phys. Rev. B {\bf 60}, 6230 (1999).

\bibitem{HoppeXTal}
R. Hoppe and H. Rieck, \textit{Die Kristallstruktur von Li$_2$CuO$_2$}, Z. anorg. allg. Chem. {\bf 379}, 157 (1970).

\bibitem{NeudertXAS}
R. Neudert, H. Rosner, S.-L. Drechsler, M. Kielwein, M. Sing, Z. Hu, M. Knupfer, M.S. Golden, J. Fink, N. Nucker, M. Merz, S. Schuppler, N. Motoyama, H. Eisaki, S. Uchida, M. Domke and G. Kaindl, \textit{Unoccupied electronic structure of Li$_2$CuO$_2$}, Phys. Rev. B. {\bf 60}, 13413 (1999).

\bibitem{RIXSReview}
A. Kotani and S. Shin, \textit{Resonant inelastic x-ray scattering spectra for electrons in solids}, Rev. Mod. Phys. {\bf 73}, 203 (2001).

\bibitem{WehtDFT}
R. Weht and W.E. Pickett, \textit{Extended moment formation and second neighbor coupling in Li$_2$CuO$_2$}, Phys. Rev. Lett. {\bf 81}, 2502 (1998).

\bibitem{HuangDDcalcs}
H.-Y. Huang, N.A. Bogdanov, L. Siurakshina, P. Fulde, J. van den Brink and L. Hozoi, \textit{Ab initio calculation of d-d excitations in quasi-one-dimensional Cu d$^9$ correlated materials}, Phys. Rev. B {\bf 84}, 235125 (2011).

\bibitem{schmitt2009}
M. Schmitt, J. M{\'a}lek, S.-L. Drechsler, and H. Rosner, \textit{Electronic structure and magnetic properties of Li$_2$ZrCuO$_4$: A spin-$1/2$ Heisenberg system close to a quantum critical point}, Phys. Rev. B {\bf 80}, 205111 (2009)

\bibitem{wolter2012}
A.U.B. Wolter, F. Lipps, M. Sch\"apers, S.-L. Drechsler, S. Nishimoto, R. Vogel, V. Kataev, B. B\"uchner, H. Rosner, M. Schmitt, M. Uhlarz, Y. Skourski, J. Wosnitza, S. S\"ullow, and K. C. Rule, \textit{Magnetic properties and exchange integrals of the frustrated chain cuprate linarite PbCuSO$_4$(OH)$_2$}, Phys. Rev. B {\bf 85}, 014407 (2012)

\bibitem{OhtaCT}
Y. Ohta, T. Tohyama and S. Maekawa, \textit{Apex oxygen and critical temperature in copper oxide superconductors: Universal correlation with the stability of local singlets}, Phys. Rev. B {\bf 43}, 2968 (1991).

\bibitem{JohnstonSub}
S. Johnston, C. Monney, V. Bisogni, K.-J. Zhou, Ro. Kraus, G. Behr, V. N. Strocov, J. M{\'a}lek, S.-L. Drechsler, J. Geck, T. Schmitt, J. van den Brink, \textit{Electron-lattice interactions strongly renormalize the charge transfer energy in the spin-chain cuprate Li$_2$CuO$_2$}. Nature Commun. {\bf 7}, 10653 (2016).



\end{thebibliography}
\end{document}